\documentclass[journal]{IEEEtran}
\ifCLASSINFOpdf
  \usepackage[pdftex]{graphicx}
  \graphicspath{{../pdf/}{../jpeg/}}
  \DeclareGraphicsExtensions{.pdf,.jpeg,.png}
\else
\fi
\usepackage{algorithmic}
\ifCLASSOPTIONcompsoc
  \usepackage[caption=false,font=normalsize,labelfont=sf,textfont=sf]{subfig}
\else
  \usepackage[caption=false,font=footnotesize]{subfig}
\fi
\usepackage{hyperref}

\usepackage{verbatim}
\usepackage{multirow}
\usepackage{booktabs}

\begin{document}
%
\title{Automating Gamification Personalization: \\To the User and Beyond}
%
%
%

\author{Luiz Rodrigues, Armando M. Toda, Wilk Oliveira, Paula T. Palomino, Julita Vassileva, Seiji Isotani%
\thanks{L. Rodrigues, A. M. Toda, W. Oliveira, P. T. Palomino, and S. Isotani are with the Laboratory of Applied Computing to Education and Advanced Social Technology, Institute of Mathematics and Computer Science, University of São Paulo, São Carlos, Brazil. E-mail: \{lalrodrigues, armando.toda, wilk.oliveira, paulatpalomino\}@usp.br, sisotani@icmc.usp.br. J. Vassileva is with the Multi-User Adaptive Distributed Mobile and Ubiquitous Computing (MADMUC) Lab, Department of Computer Science, University of Saskatchewan, Saskatoon, Canada. E-mail: jiv@cs.usask.ca.}
}

%
%

\markboth{Preprint submitted to IEEE transactions on learning technologies}%
{Rodrigues \MakeLowercase{\textit{et al.}}: ?}
%



\maketitle

\begin{abstract}
Personalized gamification explores knowledge about the users to tailor gamification designs to improve one-size-fits-all gamification.
The tailoring process should simultaneously consider user and contextual characteristics (e.g., activity to be done and geographic location), which leads to several occasions to tailor. Consequently, tools for automating gamification personalization are needed. The problems that emerge are that which of those characteristics are relevant and how to do such tailoring are open questions, and that the required automating tools are lacking.
We tackled these problems in two steps. First, we conducted an exploratory study, collecting participants' opinions on the game elements they consider the most useful for different learning activity types (LAT) via survey. Then, we modeled opinions through conditional decision trees to address the aforementioned tailoring process. Second, as a product from the first step, we implemented a recommender system that suggests personalized gamification designs (which game elements to use), addressing the problem of automating gamification personalization.
Our findings i) present empirical evidence that LAT, geographic locations, and other user characteristics affect users' preferences, ii) enable defining gamification designs tailored to user and contextual features simultaneously, and iii) provide technological aid for those interested in designing personalized gamification.
The main implications are that demographics, game-related characteristics, geographic location, and LAT to be done, as well as the interaction between different kinds of information (user and contextual characteristics), should be considered in defining gamification designs and that personalizing gamification designs can be improved with aid from our recommender system.
\end{abstract}


%
\IEEEpeerreviewmaketitle


\section{Introduction}
\label{sec:intro}

\IEEEPARstart{T}{o} improve learning technologies capability of engaging and motivating users, practitioners and researchers started to employ gamification: the use of game elements in contexts that are not games \cite{kapp2012gamification,deterding2011game,dichev2017gamifying}. Overall results from these applications are positive, showing improvements in learning outcomes such as academic achievement, conceptual and application-oriented knowledge, and motivation to learn \cite{Sailer2019}. However, there are situations in which gamification is ineffective in impacting learning outcomes, or even negative \cite{toda2018darkside}, often due to poorly designed gamification \cite{loughrey2018we,morschheuser2018design}, such as assuming that the same choices will work for all users, the one-size-fits-all approach \cite{orji2018personalizing,liu2017toward}. To overcome such failures, researchers started to investigate personalized gamification \cite{oliveira2019tailored}.

Personalized gamification concerns exploring knowledge about the users to enable providers (e.g., an instructor or the system itself) to offer game elements tailored to those users \cite{tondello2019tese}. For instance, a case would be a system or instructor changing from game elements set A to game elements set B when users are females because the latter is more tailored to these users. The premise for personalizing gamification emerged from discussions that people with, for instance, different demographic characteristics and cultural background have distinct preferences \cite{yee2016gaming,orji2017comparison}, behaviors \cite{rodrigues2018improving}, and are motivated differently \cite{orji2014modeling} and, consequently, might experience and respond to the same conditions in distinct ways \cite{knutas2019process,rodrigues2019playing}. The common practice for gamification is selecting which game elements to add to the system from a list of available elements \cite{koivisto2019rise,tondello2017elements}. Accordingly, researchers invested in providing recommendations indicating which game elements suit better users of different groups to provide personalized gamification, predominantly based on their preferences (e.g., \cite{orji2014modeling,oliveira2019tailored,baldeon2016lega}).

Mainly, those recommendations are based on users' behavioral profiles \cite{hallifax2019adaptive,tondello2016gamification,nacke2014brainhex}. However, the application context is relevant for gamification's success as well \cite{koivisto2019rise,hallifax2019factors}, and gamification designs should be aligned to it \cite{liu2017toward}. Furthermore, multiple factors (e.g., users' demographics \cite{mora2019quest,buckley2017individualising,borges2017selecting} and the system's context \cite{morschheuser2018design,Nicholson2012AUT}) moderate users' experience, either positive or negatively, but tailoring approaches often consider a single one (e.g., \cite{oliveira2020does,monterrat2017adaptation,roosta2016personalization}). These limitations reflect current gaps in the field of personalized gamification, highlighted in recent literature reviews \cite{klock2020tailored,hallifax2019adaptive}; the facts that i) personalization models should consider more than users' characteristics, such as encompassing the learning activities and geographic locations, and that ii) personalization methods should consider multiple aspects simultaneously, as well as their interactions.

To address these gaps, we sought to understand how to tailor gamified systems to the education domain by considering the learning activity at hand, the user's characteristics, and the geographic location simultaneously, as well as the interactions between all aspects taken into account. To achieve that goal, we performed an exploratory, survey-based research to capture users' preferences, a methodology that has been widely accepted and adopted by related research, as personalization is often based on user preference \cite{tondello2017elements,orji2018personalizing,oliveira2019tailored,bovermann2020towards}. As this development process is concerned with understanding which aspects (i.e., among learning activity at hand, user's characteristics, and the geographic location) affect user preference, as well as the most suitable game elements for each aspects combination, we sought to answer the following research questions:

\begin{itemize}
    \item \textbf{RQ1}: \textit{Does users' preferences differ depending on (a) their characteristics, (b) geographic location, and (c) the type of the learning activity to be performed?};
    
    \item \textbf{RQ2}: \textit{What is the most useful game elements set, from users' preferences, according to their characteristics, geographic location, and Learning Activity Type (LAT)\footnote{In the scope of this study, a LAT is defined based on its main expected outcome (see Section \ref{sec:relatedwork} for further details)}?}
\end{itemize}

RQ1 and RQ2 are interrelated as the first informs the second in terms of which characteristics should be considered when defining the most useful game elements set. Then, the challenge that emerges is that interactions from multiple characteristics lead to several combinations; for instance, five binary characteristics would lead to 25 combinations, thus, 25 recommendations, while the number of recommendations for five three-valued characteristics would exponentially increase. Consequently, it becomes of utmost importance providing a way to automate such recommendations, which corroborates another challenge of personalized gamification: automating the personalization process \cite{klock2020tailored}. Therefore, as a product of our answers, we implement a Recommender System (RS) for personalized gamification \cite{tondello2017recommender} that can be used to be informed on the most useful game elements set, according to users preferences, given an input of user's characteristics and their geographic location along with the LAT to be performed. Thus, our main contributions are the following:

\begin{enumerate}
    \item Evidence, from users' preferences, that can be used to inform researchers and practitioners on how to tailor Gamified Educational Systems (GES) to LAT, geographic location, and user characteristics;
    
    \item An RS to automate gamification personalization, which performs recommendations by considering multiple aspects simultaneously (i.e., user characteristics, geographic location, and LAT), enabling the implementation of gamification designs more aligned to their preferences; and
    
    \item Demonstrating which user characteristics impacted their preferences, along with the degree of each one's influence; thus, one might decide which user characteristic to prioritize, take into account, and/or pay more attention as, for instance, moderators of gamification's effectiveness.
    
\end{enumerate}


\section{Literature Review}
\label{sec:relatedwork}

This section provides background information on the topics covered by this paper, reasons about the literature to justify research choices, and highlights the contribution our study provides to existing literature compared to similar works. 

\subsection{Game elements} 
There are many definitions and categorizations of game elements.
In the scope of this paper, we consider game elements similar to the game element definition adopted by \cite{tondello2017elements}, which are the building blocks impacting users' experience with the system, that are characteristic to gameful systems \cite{deterding2011game}, following the vocabulary used more often by similar research \cite{hallifax2019adaptive}. 

Given the numerous game elements available, it has been common practice for each study to self-select which set of those elements to use.
Based on a literature review, \cite{tondello2017elements} presented 59 general elements. In \cite{hallifax2019factors}, the authors reviewed the literature to select 12 common game elements, without considering any content game element \cite{kapp2012gamification} due to the generic nature of their research. In both studies, game elements were selected with no consideration for the domain application, according to their purposes. Differently, \cite{oliveira2019tailored} explored an element set created from gamification on education literature \cite{nah2014gamification}, which is composed by eight options. 

Given that our research focuses on a specific domain, education, this paper differs from \cite{tondello2017elements,hallifax2019factors} by exploring a taxonomy \cite{toda2019taxonomy} containing the most common game elements (N = 21) from GES. This taxonomy was created through a rigorous, systematic process, and was validated by 19 experts in the field of gamification and games, whereas \cite{oliveira2019tailored} relied on a simpler, reduced game elements set, which was created based on a literature review. Furthermore, by selecting an expert-validated taxonomy, we ensure the game elements available are well defined, avoid using elements with the same purpose but different names, and prevent possible bias from the selection process. On top of that, the selected taxonomy also provides guidance on how the elements are expected to affect users \cite{toda2019analysing}, another advantage to those using it \cite{tondello2017elements}.

\subsection{Personalized Gamification} 

Personalizing information systems is an important aspect that should be available to enhance these systems' relevance to users \cite{liu2017toward}. Within the scope of gamified systems, a common practice to achieve personalization has been to tailor the gamification design (set of game elements) to specific user's characteristics (e.g., \cite{oliveira2020does,orji2018personalizing}). In other words, gamified systems have been personalized by performing static adaptations on the game elements it features, based on pre-defined characteristics (i.e., behavioral profile), to tailor the gamification designs \cite{hallifax2019adaptive}.

A recent literature review \cite{hallifax2019adaptive} has found that information used to drive personalization are, predominantly, users' player/gamer types \cite{tondello2016gamification,nacke2014brainhex}, followed by personality \cite{mccrae1992introduction}. 
Nevertheless, it has been shown that other user characteristics, such as gaming habits \cite{denden2017investigation} and gender \cite{toda2019planning}, also impact their preference, as well as the relationship between user demographics (i.e., age and gender) and player types \cite{tondello2016gamification} suggest the impact of those aspects. In spite of that, these aspects have been rarely explored in methods for tailoring gamification designs in education \cite{toda2019approach}. This research addresses this need by introducing an approach that exploits demographic and gaming habits as information used to drive the gamified designs' tailoring.

Furthermore, the user is not the only factor to be considered when defining gamification designs. A factor that has been often discussed as relevant for gamification effectiveness \cite{koivisto2019rise,Sailer2019,mora2019quest,hallifax2019adaptive,liu2017toward,deterding2015lens}, that is rarely considered by tailoring method, is the application context (e.g., geographic location). Specifically in the context of educational systems, an aspect researchers have recently argued as relevant, and recommended to consider when tailoring gamified systems, is the learning activity \cite{rodrigues2019thinking,hallifax2019adaptive}. This is related to the recommendation that gamified designs should match the task \cite{liu2017toward} and, given that tasks of educational systems are almost ever learning activities, personalizing the gamified designs to these activities should be accomplished.

Despite that, to the best of our knowledge, there are only two approaches for personalizing gamified designs based on learning activities \cite{bovermann2020towards,baldeon2016lega}. In \cite{baldeon2016lega}, learning activities are considered based on their main expected objective, similar to this article. In \cite{bovermann2020towards}, the learning activities are activities from Moodle (e.g., forum and quizzes). Hence, while recommendations from \cite{baldeon2016lega} can be extended to any learning activity (linked to their objective), those from \cite{bovermann2020towards} are limited to a specific set of Moodle activities. In addition, both works consider one user characteristic, personality trait and player type, respectively. Thus, they provide valuable contributions in terms of exploring learning activities, as well as presenting recommendations that consider the interaction between those and a user characteristic (e.g., player type X, learning activity Y). 

However, these studies fall into the category of methods that rely on the most often researched user characteristic, a single user characteristic is considered in each one, and the guideline from \cite{bovermann2020towards} cannot be generalized to any learning activity. Therefore, the main advances of this article compared to those works are: i) considering multiple user characteristics rarely explored simultaneously, ii) taking the context into account via learning activities and users' geographic location, and iii) providing recommendations that consider the interaction between all of those aspects that are relevant for users.

\subsection{Learning Tasks} 

To generally describe a task, one might rely on its desirable outcomes, behavioral requirements, and/or complexity \cite{wood1986task,liu2017toward}. Similarly, from the human-computer interaction perspective, a task refers to the activities required to achieve a specific goal \cite{diaper2003handbook}. Consequently, given the context of our study, a learning task refers to a set of activities that aim at some educational outcome. From this definition, it is possible to note that numerous tasks might be found in GES, which makes it infeasible to develop a specific personalization approach for each one. An alternative to that limitation is categorizing the activities, which can substantially reduce their quantity; consequently, enabling the recommendation of gamification designs to each category. 

To overcome the numerous learning tasks and categorize them, we opted to rely on the revision of Bloom's taxonomy of educational objectives \cite{krathwohl2002revision}, an approach that contributes to the learning process by matching the educational activities' gamification designs to a cognitive taxonomy \cite{baldeon2016lega}. Although there are other options available, the revision of Bloom's taxonomy is a widely cited, well-accepted taxonomy, similar to its original version \cite{bloom1956taxonomy}. It acts as a framework that can be used to classify what is expected from an educational activity (outcome), as well as its complexity \cite{krathwohl2002revision}. The revised version is composed of two dimensions: knowledge (concerned with what is to be learned; e.g., the subject of matter) and cognitive process (concerned with actions associated with learning; e.g., how to learn) \cite{krathwohl2002revision}. 

In the scope of this research, we consider the second dimension, similar to related work \cite{baldeon2016lega}. By categorizing learning activities based on the cognitive domain of such a taxonomy, we avoid having the gamification focused on the activity itself (e.g., completing a quiz or answering a forum) and allow it to be aligned with the activity's expected learning outcome, addressing the recommendation that gamification should match the task \cite{liu2017toward}. Moreover, as many GES feature tasks of varied subjects, the second dimension choice makes the approach subject-independent, focusing the gamification designs' tailoring on the activities' particular objectives while allowing it to be used regardless of the system's educational topic.

The structure of the cognitive process dimension is split into six categories: remember, understand, apply, analyze, evaluate, and create. Here, we consider each dimension a different LAT, wherein their complexity increases following the order in which they were introduced (i.e., remember is the less complex and create is the most complex). Hereafter, we refer to those as LAT1 to LAT6, also following the introduced order. Furthermore, although an activity might fit in more than one LAT, our approach considers every activity will have a predominant, main objective to be achieved. Hence, the personalization process should be based on that main goal. It is worth noting that those LAT might be split again, however, we opted to work with the high-level abstraction given that the similarities within these sub-categories might be even higher. Thus, this paper contributes a proposal that is based on the six high-level types of cognitive processes established in \cite{krathwohl2002revision}, that aids in tailoring gamification designs to different LAT, according to their predominant goal.

\subsection{Recommender Systems for Personalized Gamification}

An RS can be seen as a technique, or software tool, able to recommend items to users \cite{ricci2011introduction}. Such systems are especially valuable for cases in which several options are available, alleviating the burden of human selection by providing recommendations, often based on what other people recommend. Common applications of such systems are e-commerce, movies, and music. Recently, the use of RS has been suggested for personalized gamification \cite{tondello2017recommender}, which corroborates to our research in terms of, for instance, reducing the burden of selecting the most suitable game elements for several combinations of user characteristics, geographic location, and LAT. Next, we provide a brief overview of RS for personalized gamification following the framework by \cite{tondello2017recommender}. 

RS have three main elements: inputs, outputs, and process. 
Inputs concerns all the aspects that are received by the RS to be taken into account before doing the recommendations. There are four main types of input: user profile (e.g., demographics, personality, behavioral profile), items (e.g., game elements), transactions (e.g., the relationship between users and items; using or preferring a game element), and context (e.g., geographic location, activities to be done). 
Outputs are ratings related to the choices that the RS made from the input received. For instance, if items are game elements, the output would be the rating of each one.
The process is the core part of the RS, concerning the method through which it will perform the recommendations. There also are four main recommendation methods. Content-based recommenders are based on knowledge of the application, such as data log, or empirical and theoretical information. The collaborative filtering method exclusively depends on data collected implicitly or explicitly from interactions with a system. Context-aware recommenders are those that explore information of the context to make their choices. Lastly, hybrid recommenders aim at using two or more of the previous approaches together.

Generally, there is a lack of technological support for gamifying educational environments \cite{dicheva2018motivational}. Accordingly, the literature on personalized gamification lacks concrete RS implementations, demonstrate by recent literature reviews finding only four studies that relate to RS or other forms of automating gamification personalization. Among those, one is the framework proposal itself \cite{tondello2017recommender}, whereas the remaining are theoretical/conceptual models with no concrete implementations available for third-parties use \cite{knutas2019process,xu2015based,monterrat2014toward}. Differently, we present and provide an RS for personalizing gamification, which was built upon findings from the study of this article. Hence, we advance the literature with a free, hybrid RS as it uses both contextual as well as empirical information from users' preferences.

\subsection{Summary}

Table \ref{tab:rw_comparison} summarizes and demonstrates the points in which this study differs from related works based on the discussion previously presented. 
As shown, most studies focus on user characteristics, few consider the task to be done, and none but this one take into account geographic locations. Additionally, the few works that consider information from the user and the task provide recommendations based on two factors (one from each kind). On the other hand, our approach was developed considering nine aspects, of which eight were found to be significant (see Section \ref{sec:results}) and, therefore, are considered in the product from our research (see Section \ref{sec:rs}). This final product is another key difference. Whereas previous research only provides conceptual/visual guidelines, this study contributes with technological aid for the designing of personalized gamified systems. This also differs from research on recommender systems for gamification \cite{knutas2019process,xu2015based,monterrat2014toward} as those provide no concrete implementations from their proposals. 

\begin{table}[htb]
\centering
\caption{Summary comparison of related works.}
\label{tab:rw_comparison}
\begin{tabular}{lcccccc}
\hline
      & \multicolumn{4}{l}{Recommends game elements based on} & \\
\hline
Study                           & User & Task & GL      & N Factors    & Product \\
\hline
\cite{tondello2017elements}     & X    &      &         &              & Conceptual \\
\cite{hallifax2019factors}      & X    &      &         &              & Conceptual \\
\cite{oliveira2019tailored}     & X    &      &         &              & Conceptual \\
\cite{tondello2016gamification} & X    &      &         &              & Conceptual \\
\cite{denden2017investigation}  & X    &      &         &              & Conceptual \\
\cite{toda2019planning}         & X    &      &         &              & Conceptual \\
\cite{bovermann2020towards}     & X    & X    &         & 2            & Conceptual \\
\cite{baldeon2016lega}          & X    & X    &         & 2            & Conceptual \\
This                            & X    & X    & X       & 8            & Technology \\
\hline
\multicolumn{6}{l}{GL = Geographic location.}
\end{tabular}
\end{table}

\section{Study}
\label{sec:study}

The goal of this research is to understand how to tailor GES to LAT, geographic location, and users' characteristic. To achieve that goal, we addressed two research questions, as outlined in Section \ref{sec:intro}. To answer them, we performed a survey-based research asking participants to indicate their preferred game elements for each LAT. Up to date, this methodology is the most used by similar works \cite{klock2020tailored} and has been widely accepted given the number of related research following it \cite{tondello2017elements,orji2018personalizing,oliveira2019tailored,bovermann2020towards}. Therefore, we considered it the most adequate approach to adopt. This study also follows an exploratory approach, which aims to understand possible relations between the observable variables, in order to create possible research guidance \cite{lazar2017research}. This section presents an overview of this study development process, as well as further describes the material and methods followed.

\subsection{Overview}

In developing this research, three factors had to be defined: what domain, how to interpret the tasks, and which user characteristic to consider. \textbf{First}, we opted for the education domain, which is the one gamification research has focused the most \cite{koivisto2019rise} and, both positive \cite{Sailer2019} and negative \cite{toda2018darkside} outcomes have been found, showing the need for further research. \textbf{Second}, given the domain, users will perform learning activities when using the gamified systems; as one might create numerous of those activities, our approach considers activities types, based on the revised Bloom's taxonomy \cite{krathwohl2002revision}, an established, well-accepted taxonomy within the educational context. \textbf{Third}, we chose to focus on users' demographic characteristics and gaming habits and preferences, deepening into aspects that have been discussed as relevant factors \cite{mora2019quest,van2018need,denden2017investigation} but received less attention from the academic community compared to the most used ones \cite{hallifax2019factors}.

Then, to achieve the desired understanding, we developed Conditional Decision Trees (CDT), which takes into account the interactions between all input variables to determine the most suitable game element for that input set. During data collection, we operationalized gamification designs as the top three game elements subjects prefer the most, provided game elements (N = 21) extracted from the expert-validated taxonomy by \cite{toda2019taxonomy}, and operationalized LAT as the six cognitive process types defined in \cite{krathwohl2002revision}. As we considered gamification designs to be composed of participants' top three game elements, ranked by their preferences. 


\subsection{Procedure}

The following five steps were performed to develop our approach for tailoring gamified designs to LAT and users. 

\begin{enumerate}
    \item \textbf{Survey development}: defining the survey design and sections and the game elements and LAT to consider;
    
    \item \textbf{Data collection}: disclosing the survey online, through Amazon's Mechanical Turk\footnote{https://www.mturk.com/} (MTurk), to collect participants opinions.
    
    \item \textbf{Data analysis}: running analyses to identify which characteristics impact users' preferences.
    
    \item \textbf{Users preferences analysis}: investigating our findings to identify how to tailor educational systems' gamification designs to users, geographic location, and LAT.
    
    \item \textbf{RS design}: developing a free, ready-to-use resource, based on our findings, to aid those who want to tailor their educational systems' gamified designs.
    
\end{enumerate}


\subsection{Survey} 

The survey was developed online\footnote{Online survey: http://bit.ly/2JWxwqs} and can be viewed in the appendix. Its design was defined in four steps. First, two researchers brainstormed and developed an initial version. Second, three other researchers revised it and provided feedback on how to improve it. Following, the survey was improved accordingly and, lastly, we ran a pilot study with 50 participants.

The final version has four sections: consent form, demographics, gaming background, and preferences. 
In the \textbf{consent form}, all respondents were informed to be participating in a research and agreed all information provided would be used to research ends only. The \textbf{demographics} and \textbf{gaming background} captured participants' gender, age, living country, highest level of education, and MTurk identifier to avoid repeated completions and for how many years the participants researched/worked with gamification (0 for those who did not), how much time (in hours) they spend with games per week, and their preferred game genre and playing setting, respectively. Lastly, in the \textbf{preferences} section, participants ranked the top three game elements they prefer the most when performing each of the six LAT.

The 21 game elements available were: Acknowledgment; Chance; Competition; Cooperation; Economy; Imposed Choice; Level; Narrative; Novelty; Objectives; Point; Progression; Puzzles; Rarity; Renovation; Reputation; Sensation; Social Pressure; Stats; Storytelling; Time Pressure. Further descriptions of these elements can be seen in the study provided by \cite{toda2019taxonomy}. The LAT are those introduced in Section \ref{sec:relatedwork} (remember, understand, apply, analyze, evaluate, and create). For further information about each one, see \cite{krathwohl2002revision}. Thus, the last section had seven items, one for each LAT and a repeated item to assess participant attention/consistency (see next section). 

Each of those seven items had three sub-items, allowing the participant to select the rank-one, -two, and -three game elements, in which the 21 game elements were possible answers. Nevertheless, the same game element could not be selected twice within the same item, that is, each participant's top three should be composed of three different game elements. A sample question was \textit{Indicate the three gamification elements you consider will help you the most when performing an activity you need to REMEMBER something (e.g., remember what the ‘+’ symbol means in arithmetic operations).}, whereas other items of the same section differed only in the LAT (e.g., understand instead o remember) and the example at the end of the item. All items had basic mathematical examples due to the generality of the topic.

Additionally, we highlight that this top-three survey design was adopted due to the number of both game elements (21) and LAT (six), which would lead to a questionnaire with 126 items if subjects should, similar to related work \cite{tondello2017elements,oliveira2019tailored}, provide a rating for each gamification element through a Likert-scale. That is, participants would answer to 21 items six times; one time per LAT. Thus, we opted for one item per LAT, featuring three options each, to reduce effort, tiredness, and time spent in completing the survey aiming to improve answers' reliability. Lastly, note that the survey sections' order was fixed (the same as previously introduced) but, within each section, the items' order was randomized.

\subsection{Data Collection and Filtering} 

We recruited participants through crowdsourcing (MTurk). We made this choice to increase our sample size, similarly to related research (e.g., \cite{hallifax2019factors,tondello2017elements}), an approach that has been recommended in the literature \cite{casler2013separate,landers2015inconvenient} to improve external reliability \cite{wohlin2012}. No participant restriction was enforced to avoid selection biases and everyone who completed the survey received a fixed remuneration. 

Nevertheless, recent studies (e.g., \cite{hallifax2019factors,tondello2017elements}) have employed additional items to the survey's long sections to assess whether participants are paying attention and providing consistent answers. Based on those specific items' answers, researchers filter participants according to some assertion threshold (e.g., discarding those who failed in more than one item \cite{hallifax2019factors}). In this study, we adopted a similar approach. On the \textit{preferences} section, we added a repeated question for one LAT, which allowed us to assess whether the participant was consistent in his or her answer (i.e., did they select the same top-three game elements in both items?). Participants' remuneration was not conditional to consistently answering, neither participants were warned about the repeated item, aiming to improve the reliability assessment.

Following related work, we adopted a tolerance for inconsistent completions. Hence, discarding all participants that provided consistent answers in less than two out of the three game elements. For instance, one selected Acknowledgment, Chance, and Competition and, then, in the repeated question, selected Acknowledgment, Cooperation, and Economy. This participant would be discarded by selecting two different game elements for the same question.
In total, 1018 individuals have completed the survey, from which 657 answers were discarded based on our criteria. Thus, the final dataset contains 361 consistent answers. The description of these reliable, valid answers is shown in Table \ref{tab:data_desc}. 

Overall, our sample is composed of adults (51.5\% males, 47.4\% females, and 1.1\% others) with 32 years on average ($\pm$11) and undergraduate or higher degrees (65.4\%). Hence, we might expect our sample to feature responsible people with good educational background. Furthermore, despite the large majority never researched gamification (91\%), there is an interesting variation in their preferred playing setting (Singleplayer: 59\%; Multiplayer: 41\%) as well as game genre (20\% for the most preferred genre: Role Playing Game), with an overall playing time of 12 hours ($\pm$13) per week. Thereby, we might expect participants to be familiar with games and their elements. 

\begin{table}[htb]
\centering
\caption{Dataset description.}
\label{tab:data_desc}
\begin{tabular}{p{2.2cm}rlr}
\hline
Value & N(\%) & Value & N(\%) \\
\hline
\multicolumn{2}{c}{\textit{Gender}}                      & \multicolumn{2}{c}{\textit{Preferred game genre}}        \\
Female                                     & 186 (0.515) & Role Playing Game                          & 75 (0.208)  \\
Male                                       & 171 (0.474) & Adventure                                  & 61 (0.169)  \\
Other Gender                               &   4 (0.011) & Action                                     & 60 (0.166)  \\
\multicolumn{2}{c}{\textit{Country}}                     & Strategy                                   & 50 (0.139)  \\
United States                              & 259 (0.717) & Other Genre                                & 115(0.319)  \\
India                                      &  22 (0.061) & \multicolumn{2}{c}{\textit{Preferred playing setting}}   \\
United Kingdom                             &  20 (0.055) & Singleplayer                               & 214(0.593)  \\
Canada                                     &  18 (0.050) & Multiplayer                                & 147(0.407)  \\
Brazil                                     &  16 (0.044) & \multicolumn{2}{c}{\textit{Researched gamification}}     \\
Italy                                      &   6 (0.017) & No                                         & 329 (0.911) \\
Germany                                    &   5 (0.014) & Yes                                        & 32  (0.089) \\
Spain                                      &   3 (0.008) & \multicolumn{2}{c}{\textit{Age}}\\
Australia                                  &   2 (0.006) & Mean & 32.615 \\
Netherlands                                &   1 (0.003) & SD & 11.299 \\
Albania                                    &   1 (0.003) & Min. & 18.000 \\
France                                     &   1 (0.003) & 25\% & 24.000 \\
Ireland                                    &   1 (0.003) & 50\% & 29.000 \\
Poland                                     &   1 (0.003) & 75\% & 39.000 \\
Turkey                                     &   1 (0.003) & Max. & 75.000 \\
Austria                                    &   1 (0.003) & \multicolumn{2}{c}{\textit{Weekly playing time (hours)}} \\
Nigeria                                    &   1 (0.003) & Mean & 12.874 \\
Belize                                     &   1 (0.003) & SD & 13.782   \\
Jamaica                                    &   1 (0.003) & Min. & 0.000  \\
\multicolumn{2}{c}{\textit{Highest education level}}     & 25\% & 4.000  \\
Undergraduate                              & 161 (0.446) & 50\% & 10.000 \\
High School                                &  81 (0.224) & 75\% & 20.000 \\
MsC                                        &  63 (0.175) & Max. & 112.000 \\
Technical education                        &  30 (0.083) & \\
Other Education                            &  14 (0.039) & \\
Ph.D                                       &  12 (0.033) \\ 
\hline
\end{tabular}
\end{table}

\subsection{Data Analysis} %

For data analyses, we worked with the \textit{party} R package \cite{hothorn2006unbiased}. A decision tree is a classification algorithm that selects an output based on the interaction between elements from an input set \cite{safavian1991survey}. Besides handling interactions, which is a key point from our objective, the decision tree algorithm provides another three positive points that led us to choose it. 
First, it allows visualizing the rules followed to determine the output. Therefore, we can comprehensively discuss and understand how game elements are selected, given an input set (user data and LAT). 
Second, it demonstrates which aspects are more or less important, as the main ones are in the tree's top, and vice-versa. It also ignores unnecessary inputs, excluding from the three those that do not contribute. Hence, providing insights on which aspects influence users' preferences, as well as which are most influencing ones from those we studied.
Third, the algorithm itself determines how each characteristic will be split (e.g., should age be split in 18-28, 29-39 or 18-23, 24-29, 30-39?), removing human bias that are likely to be inserted in this process. 

Especially for reducing bias, we chose to use CDT rather than the traditional version of the algorithm. Traditional decision tree implementations suffer from bias in the selection process when inputs have many splitting points (e.g., how to split age, playing time, or country - when there might be more than 100 countries?), and they are likely to lead to overfitting as the algorithms do not consider whether a splitting/selection improves the tree \cite{mingers1987expert}. CDT, on the other hand, address these gaps by following a statistical approach that takes into account measures' distributions during splits and variable selection \cite{hothorn2006unbiased}. 

Thus, according to our goal, data captured via our survey (Table \ref{tab:data_desc}) was entered as input to generate three CDT. Each tree's output was users' preferences for either their first, second, or third selected game elements. Accordingly, each tree predicts a user's preferred game element. Note that the dataset described in Table \ref{tab:data_desc} is in the \textit{wide format}; that is, one row per participant, and one column for their preference on each LAT (one column for LAT1, one for column LAT2, and so on). Then, to generate trees able to distinct users' preferences from one LAT to another, we converted the dataset to \textit{long format}; that is, six rows per participant, a new column indicating the LAT each row corresponds to, and a single column indicating the preferred game element from each user for each LAT. We highlight that, although this increases the size of the dataset inputted to the CDT, the characteristics' distribution remains the same.

\section{Results}
\label{sec:results}

First, this section presents the overall answers collected from the survey. Then, it explores the CDT generated from our data, discussing the answers for our research questions in light of insights gained from them.

\subsection{Overall Survey Responses}

Table \ref{tab:overall_preference} demonstrates how much each game element was selected by the participants, independent of selecting the game element as first, second, or third choice. Overall, Acknowledgment was the most selected game element (11\%), followed by Objectives (10\%) and Cooperation (8\%). On the other hand, Reputation, Rarity, and Social Pressure are the less chosen, summing up to less than 3\% together. Nevertheless, one should note the participants' selections distribution, with the game element selected the most having only 11\% of those, and also considering that the fifth element selected the most - Narrative (7\%) - difference to the first was just four percentual points.

\begin{table}[htb]
\centering
\caption{Overall participants' preferences.}
\label{tab:overall_preference}
\begin{tabular}{lrlr}
\hline
Game element    & N(\%)       & Game element    & N(\%)  \\
\hline
Acknowledgment  & 740 (0.114) & Storytelling    & 290 (0.045) \\
Objectives      & 654 (0.101) & Stats           & 281 (0.043) \\
Cooperation     & 545 (0.084) & Point           & 213 (0.033) \\
Competition     & 507 (0.078) & Time Pressure   & 180 (0.028) \\
Narrative       & 420 (0.065) & Sensation       & 126 (0.019) \\
Progression     & 405 (0.062) & Renovation      & 117 (0.018) \\
Level           & 381 (0.059) & Novelty         & 106 (0.016) \\
Imposed Choice  & 363 (0.056) & Reputation      &  76 (0.012) \\
Puzzles         & 352 (0.054) & Rarity          &  64 (0.010) \\
Economy         & 327 (0.050) & Social pressure &  37 (0.006) \\
Chance          & 314 (0.048) & \\
\hline
\end{tabular}
\end{table}

More specifically, Table \ref{tab:preference_per_selection} shows how much each game element was chosen as the first, second, and third options. It demonstrates Acknowledgment was selected the most as first and Objectives as both the second and third choices. Again, the proximity in preferences must be noted, with the most selected game element being chosen only 20\% of the times. These results suggest the variety of participants' preferences, further making the case for understanding how these preferences change according to users' characteristics as well as the LAT they expect to perform and their geographic location.

\begin{table}[htb]
\centering
\caption{Participants' preferences for first, second, and third game element. Data represented as N(\%).}
\label{tab:preference_per_selection}
\begin{tabular}{lrrr}
\hline
Game element       & \multicolumn{1}{c}{First} & \multicolumn{1}{c}{Second} & \multicolumn{1}{c}{Third}  \\
\hline
Acknowledgment    & \textbf{426 (0.197)} & 130 (0.060) & 184 (0.085) \\
Chance             & 109 (0.050) & 105 (0.048) & 100 (0.046) \\
Competition        & 211 (0.097) & 158 (0.073) & 138 (0.064) \\
Cooperation        & 176 (0.081) & 197 (0.091) & 172 (0.079) \\
Economy            &  81 (0.037) & 116 (0.054) & 130 (0.060) \\
Imposed Choice     & 155 (0.072) & 116 (0.054) &  92 (0.042) \\
Level              & 104 (0.048) & 141 (0.065) & 136 (0.063) \\
Narrative          & 162 (0.075) & 168 (0.078) &  90 (0.042) \\
Novelty            &  19 (0.009) &  44 (0.020) &  43 (0.020) \\
Objectives         & 191 (0.088) & \textbf{262 (0.121)} & \textbf{201 (0.093)} \\
Point              &  37 (0.017) &  80 (0.037) &  96 (0.044) \\
Progression        &  57 (0.026) & 154 (0.071) & 194 (0.090) \\
Puzzles            & 138 (0.064) & 116 (0.054) &  98 (0.045) \\
Rarity             &  13 (0.006) &  22 (0.010) &  29 (0.013) \\
Renovation         &  28 (0.013) &  41 (0.019) &  48 (0.022) \\
Reputation         &   8 (0.004) &  30 (0.014) &  38 (0.018) \\
Sensation          &  23 (0.011) &  49 (0.023) &  54 (0.025) \\
Social pressure    &   9 (0.004) &  11 (0.005) &  17 (0.008) \\
Stats              &  77 (0.036) &  87 (0.040) & 117 (0.054) \\
Storytelling       & 107 (0.049) &  84 (0.039) &  99 (0.046) \\
Time Pressure      &  35 (0.016) &  55 (0.025) &  90 (0.042) \\
\hline
\end{tabular}
\end{table}

\subsection{Conditional Decision Trees Overview}

\begin{figure*}
\centering
\includegraphics[width=\textwidth]{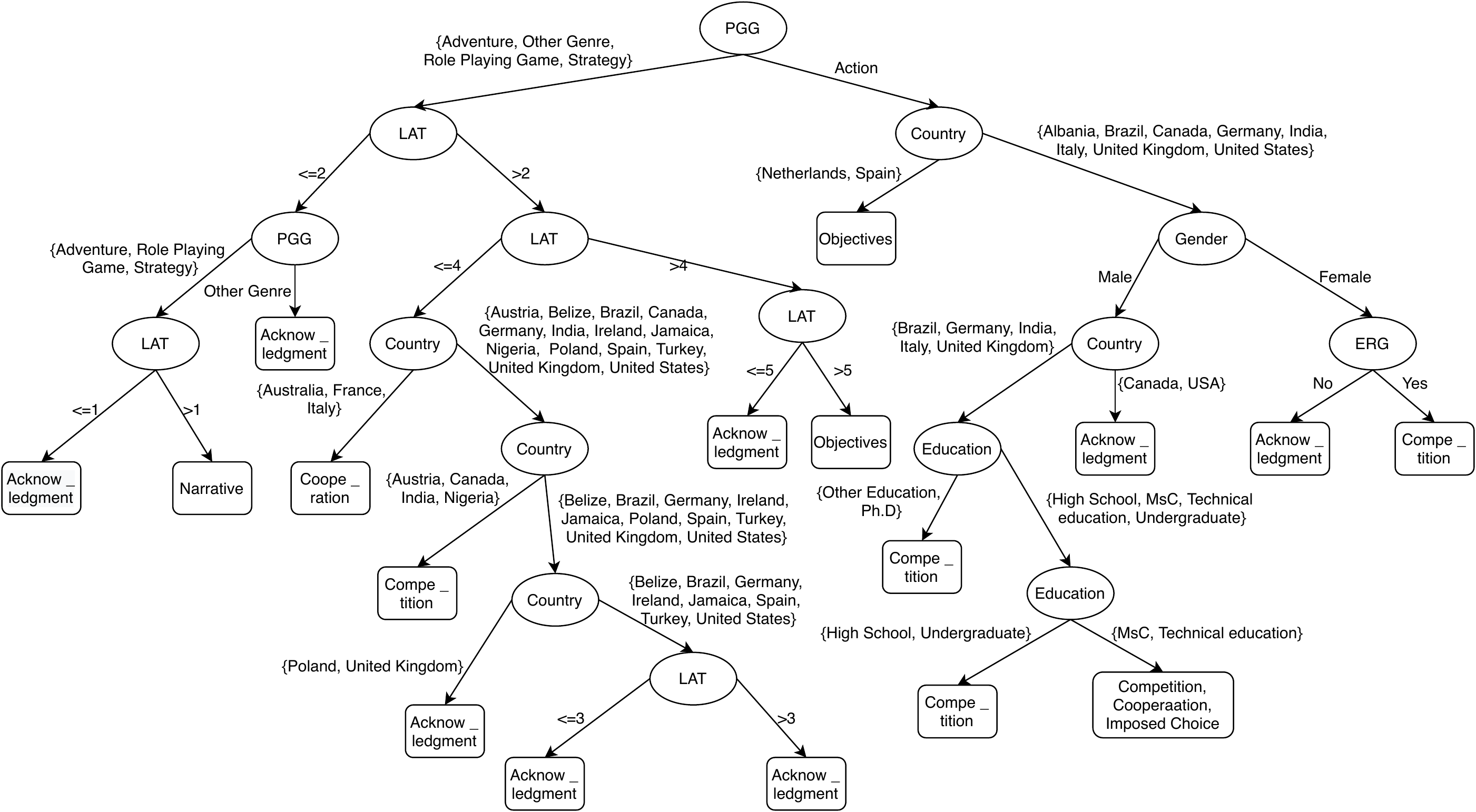}
\caption{Conditional decision tree for participants most preferred game element. Codes refer to preferred game genre (PGG), learning activity type (LAT), and experience researching gamification (ERG).}
\label{fig:cdt1}
\end{figure*}

Concerning participants' number one choice, that is, the game element they prefer the most for each LAT, our first CDT - CDT1 - was constructed. Similarly, our second and third CDT - CDT2 and CDT3, respectively - concern the game element participants selected as the second- and third-preferred ones for each LAT. CDT1 is shown in Figure \ref{fig:cdt1}, where circles represent decision nodes and rectangles are leaf ones. Decision nodes function as if/else statements. For instance, the first node tests if, for a given input, the preferred game genre is equal to adventure, other genres, role playing game, or strategy (left), or equal to action (right). Based on the answer, it is decided whether one should follow to the left or right path of the tree. This procedure is iteratively repeated for each decision node until reaching a leaf node. Leaf nodes indicate the tree's output, which is the preferred game element in our case. Hence, for someone who preferred the game genre is action and lives in the Netherlands, CDT1 would recommend the Objectives game element.

\begin{figure}
\centering
\includegraphics[width=0.49\textwidth]{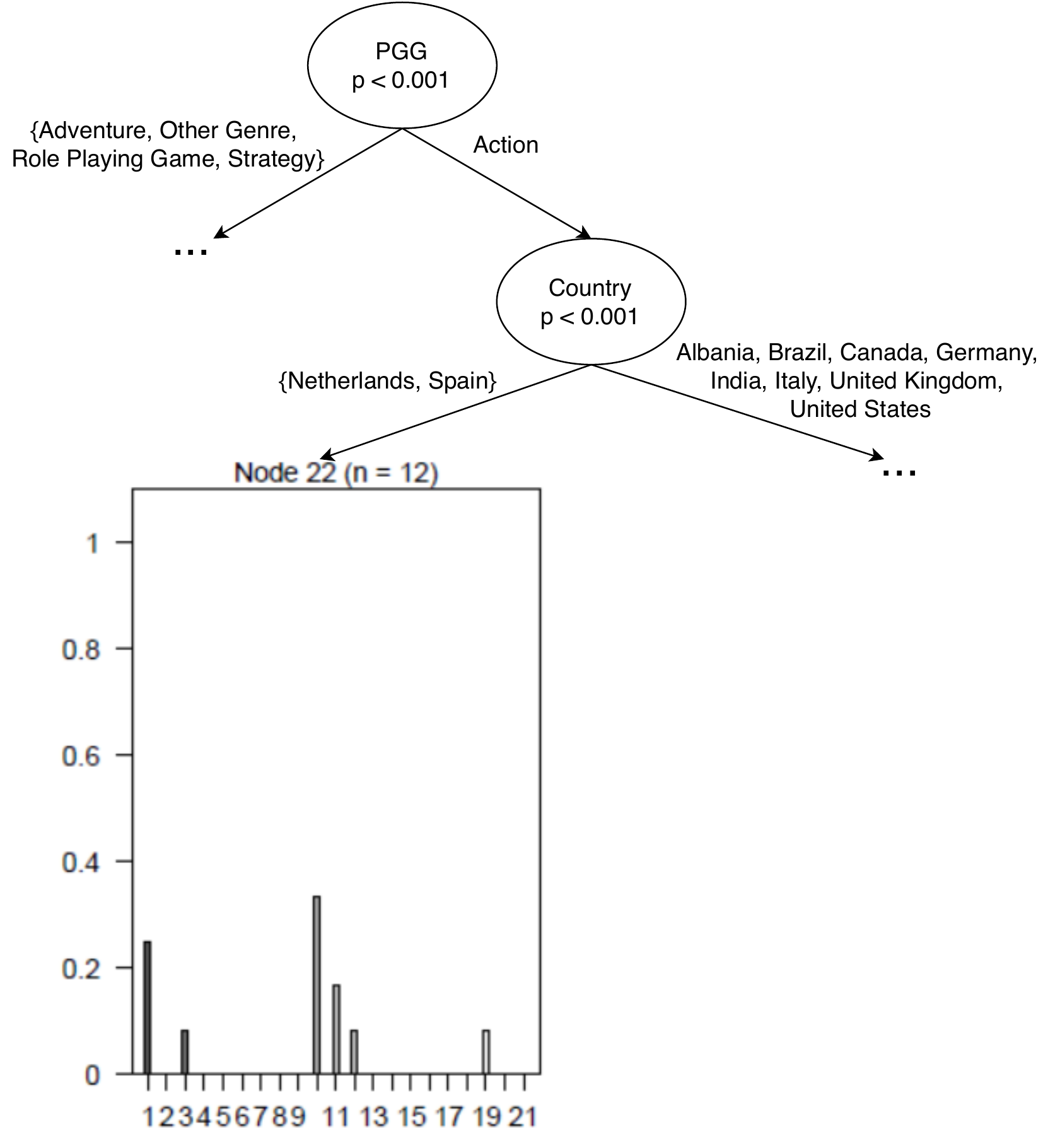}
\caption{Partial visualization of a CDT generated from our data (CDT1), illustrating how its leaf nodes demonstrate the distribution from participants' selections.}
\label{fig:cdt1_subtree_original}
\end{figure}

One should note, however, that the tree in Figure \ref{fig:cdt1} is a simplified version compared to the original tree generated from the R package \textit{party}. The original version has two main differences. First, it shows p values from each split, demonstrating they are significant splits. Second, their leaf nodes present bar plots, demonstrating to what extent each game element was selected for that path's sample. For instance, for those who prefer the game genre action and live in the Netherlands, the tree allows identifying which game elements this sub-sample selected, as well as the percentage selection of each one. Figure \ref{fig:cdt1_subtree_original} partly shows the original version of CDT1, focusing on the case aforementioned. Note that game elements are numerically represented in the figure's barplot for simplicity. Each element's number is defined alphabetically, with Acknowledge represented by number one and Time pressure by number 21 (recall that the 21 game elements are, in alphabetical order, Acknowledgment, Chance, Competition, Cooperation, Economy, Imposed Choice, Level, Narrative, Novelty, Objectives, Point, Progression, Puzzles, Rarity, Renovation, Reputation, Sensation, Social Pressure, Stats, Storytelling, Time Pressure).

From Figure \ref{fig:cdt1_subtree_original}, we see the 12 participants who live in Netherlands or Spain and prefer action games chose six different game elements, with Objectives (number 10 on the x-axis) being the most selected one, followed by Acknowledgment (number one on the x-axis). This insight can be used to recommend the most preferred game element (e.g., Objective) or to provide ratings on the most likely preferred ones (e.g., Objective, Acknowledgment, and Point - number 11 on the x-axis - are the most preferred elements). Considering this context, we highlight Figure \ref{fig:cdt1} only presents the most preferred game element for the sake of limited space, as the full image would be readable within the article template. For similar reasons, CDT2 and CDT3 are not shown in the article. Nevertheless, the full images, with barpots for all leaf nodes, from all CDT we created, are available in the appendix. 



\subsection{RQ1: Characteristics that Impact User Preferences}

\textbf{RQ1} concerns finding out which aspects, among user characteristics, geographic location, and LAT, impact users' preferences for game elements. Therefore, we analyze which of those appear in our CDT to identify the ones that influence participants' preferences.

\textit{CDT1} used six of the nine (eight from Table \ref{tab:data_desc} plus LAT) inputs, namely preferred game genre, LAT, gender, country, experience researching gamification, and education, which appeared in the tree in the same order as presented here. Thus, for participants number one choice, those are the characteristics that impacted their preferences, with preferred game genre and education being the most and the less influencing ones.
\textit{CDT2} also used six out of the nine inputs; those are country, LAT, preferred game genre, gender, preferred playing setting, and weekly playing time; the order of relevance for the tree is the same as the one they were presented here. Thus, for participants number two choice, those are the six characteristics that impacted their preferences, with country and weekly playing time being the most and the less influencing ones.
\textit{CDT3} used five of the nine inputs, country, preferred game genre, experience researching gamification, LAT, and education, with the same order of relevance as presented here. Thus, these are the characteristics that influenced participants' preferences for their third choice, in which country was the most relevant one, as opposed to education and LAT that were both the less relevant ones.

Based on these findings, we might answer \textbf{RQ1} with evidence that factors impacting users' preferences are country, LAT, preferred game genre and playing setting, gender, experience researching gamification, weekly playing time, and education. In addition, we also found the order of importance of these characteristics for each of the three selections. This finding is summarized in Table \ref{tab:importance}, which demonstrates the highest level of the tree where each characteristic appears (because one might appear multiple times and at different levels). Consequently - as the higher the level, the more the importance - allowing us to identify each one's importance.

\begin{table}[htb]
\centering
\caption{Level in which each characteristic appeared in the CDT of each users' choices.}
\label{tab:importance}
\begin{tabular}{lcccccccc}
\hline
Choice &  Cnt & LAT & PGG & PPS & G & ERG & WPT & Edu \\
\hline
First  &  4   & 2   & 1   &     & 3 & 4   &     & 5   \\
Second &  1   & 2   & 2   & 5   & 5 &     & 6   &     \\
Third  &  1   & 4   & 2   &     &   & 2   &     & 4   \\
\hline
\multicolumn{9}{p{0.45\textwidth}}{Cnt = country; LAT = learning activity type; PGG = preferred game genre; PPS = preferred playing setting; G = gender; ERS = experience researching gamification; WPT = weekly playing time; Edu = education.}
\end{tabular}
\end{table}

\subsection{RQ2: Most useful Game Elements Sets from User' Preferences}
\label{sec:most_useful_game_elements}

\textbf{RQ2} concerns identifying which is the most useful game elements set, given users' characteristics, the type of the learning activity they will perform, and their geographic location, according to participants' preferences. From the three CDT we generated, it is worth mentioning that CDT1, CDT2, and CDT3 have 17, 16, and 15 terminal nodes, respectively. This means that, together, all trees provide recommendations for 48 input set combinations. Presenting a complete description of the recommendation for each of these combinations is unfeasible. Nevertheless, we demonstrate recommendations for specific cases to illustrate the most useful game elements set for such cases, according to our findings.

First, let us consider the simple case where one wants to personalize gamification to LAT only, without considering any user characteristic. To illustrate that case, we split our dataset in six: each one containing only rows of one LAT. Then, we predict the output from each of our CDT using each sub-dataset, in which the results are shown in Table \ref{tab:recommendation_lat}. There are cases (e.g., first, second, and third rows) in which the same element is recommended as second and third preferred, for instance. Although one participant could not select the same element for both cases, this corroborates the fact that the most select game element as second and third, considering the overall sample, was Objectives. Accordingly, our CDT recommend the same element as the second and third choices. With that in mind, our findings suggest that the most useful game elements set, considering LAT and no user characteristic, for LAT1 is Acknowledgment and Objectives, for LAT2 is Narrative and Objectives, and so on.

\begin{table}[htb]
\centering
\caption{Recommendations for personalizing gamification to LAT only, without considering any user characteristic, based on our dataset.}
\label{tab:recommendation_lat}
\begin{tabular}{lcccccc}
\hline
LAT & First          & Second     & Third \\
\hline
1   & Acknowledgment & Objectives & Objectives \\
2   & Narrative      & Objectives & Objectives \\
3   & Acknowledgment & Objectives & Objectives \\
4   & Acknowledgment & Objectives & Acknowledgment \\
5   & Acknowledgment & Level      & Point \\
6   & Objectives     & Objectives & Progression \\
\hline
\end{tabular}
\end{table}

Among the main contributions from our approach, is its ability to handle multiple characteristics simultaneously, as well as the interaction between these characteristics. Therefore, let us assess the cases in which one wants to personalize gamification for a learning activity wherein students need to recall some content from long-term memory (remembering, LAT1) and then perform a second activity in which they need to evaluate others' opinions (evaluate, LAT5). Additionally, let us compare the most useful game elements set for Brazilian and Americans performing such activities. For the sake of simplicity, let us assume all students are males, never researched gamification, High School degree is their highest education level, play similar amounts of time per week (10 hours), and prefer the same game genre and playing setting: action and singleplayer, respectively\footnote{This fixed combination was selected arbitrarily, aiming to simplify the illustration. Other characteristics were not mentioned as they were found not to influence user preferences (see Table \ref{tab:importance})}. In this context, the recommendations are likely to vary due to changes in LAT, as well as geographic location (country), as all other relevant characteristics are the same. Table \ref{tab:recommendation_lat_pgg} demonstrates the recommendations.

\begin{table}[htb]
\centering
\caption{Recommendations, depending on LAT and country, for an arbitrarily selected sample: males, who never researched gamification and have High School degree as their highest education level, play 10 hours per week, and prefer playing action games alone.}
\label{tab:recommendation_lat_pgg}
\begin{tabular}{lccc}
\hline
Combination         & First          & Second      & Third  \\
\hline
LAT1 - USA    & Acknowledgment & Competition & Competition   \\
LAT1 - Brazil & Competition    & Competition & Time pressure \\
LAT5 - USA    & Acknowledgment & Level       & Point         \\
LAT5 - Brazil & Competition    & Level       & Point         \\
\hline
\end{tabular}
\end{table}

As shown in Table \ref{tab:recommendation_lat_pgg}, our findings suggest the most useful game element set for LAT1 for Brazilians is Acknowledgment and Competition, whereas that for Americans is Competition and Time pressure. For LAT5, the recommendation for Brazilians is Acknowledgment, Level, and Point, while that for Americans differ by suggesting Competition rather than Acknowledgment. Hence, highlighting the impact of contextual factors on users' preference, in which that differed depending on the LAT they were expecting to perform, as well as their geographic location.

In summary, we demonstrated which are the most useful game element set for specific combinations of user and contextual characteristics. We did not show the recommendations for all combinations because our CDT have 48 leaf nodes, which means they provide recommendations for 48 combinations of the characteristics considered in this study, an unfeasible discussion for this article. Despite that, one can follow CDT1 (Figure \ref{fig:cdt1}) to determine the most useful game element (number one of three) for any combination of the characteristics we studied. Both CDT2 and CDT3 can be analyzed similarly and accessed in the appendix for finding the recommendations for users' second and third preferences. Lastly, we acknowledge that although comprehensible, the process for gathering insights from the three CDT can be improved. Our RS, which contributes to that improvement, is presented in the next section.

\section{The Recommender System}
\label{sec:rs}

To cope with the complexity of determining recommendations from visual inspection of CDT, we converted our tree CDT into an RS. This system encapsulates all trees and simplifies the task of determining which game elements to use given a user, a LAT, and a geographic location. Next, we further describe the characteristics of our RS, then, we briefly present technical concerns on how our CDT were converted into a free, easy-to-use system able to automate the personalization of GES.

We characterize our RS according to the framework for RS for personalized gamification introduced in \cite{tondello2017recommender}. Our RS considers six user inputs. Those are their preferred game genre and playing setting, weekly playing time, gender, highest education level, and whether the user researched gamification before. In addition, the user's living country, as well as the LAT that will be gamified, must also be entered, inputs related to the context \cite{khemaja2017building}. Items are the game elements users could choose in the survey, the 21 game elements from the taxonomy proposed and validated in \cite{toda2019taxonomy}. Lastly, the transactions concern users' preferred game elements (i.e., user with characteristics X, from geographic location Y, prefers element Z for LAT W), which is defined according to our findings.

The method adopted for output selection characterizes our RS as a hybrid recommender \cite{tondello2017recommender}. The input involves two contextual characteristics, geographic location and the LAT to be performed. Accordingly, the method would be characterized as a context-aware recommender. However, the selection process also relies on empirical information from our findings, which concerns a  content-based recommender. Thus, our RS is a hybrid recommender due to exploring the characteristics of two methods. Lastly, our RS outputs are the ratings for game elements, defined according to the percentual of each game element's selection for that input. Then, if we consider the case shown in Figure \ref{fig:cdt1_subtree_original}, there would be ratings for six ratings because the other 15 game elements were not chosen, and Objectives game element rating would be roughly 0.30 whereas that of Acknowledgment would be around 0.20. Furthermore, as there are tree CDT, the output will show the rating of each game element as the first-, second-, and third-preferred game element.

Table \ref{tab:ratings_sample} exemplifies an output of our RS. It demonstrates a full output of the RS with the ratings for all game elements when considered as first-, second-, and third-preference. As we discussed previously, for participants' number one preference, the game element with the highest rating is Objectives (0.333), followed by Acknowledgment (0.25) and Point (0.167). For participants number two choice, the highest ratings are for Competition (0.149) and Change (0.144), followed by Cooperation in third place (0.117). For their third preferred element, Competition holds the highest rating (0.176), followed by Acknowledgment and Cooperation (0.078 for both). When using the RS for other input sets, similar outputs will be given, likely with different ratings for each game element. Hence, based on outputs as that shown in Table \ref{tab:ratings_sample}, one can assess which elements are more likely to be the preferred ones for a given situation and define their gamification design accordingly.

\begin{table}[htb]
\centering
\caption{Ratings of our RS for people who live in Netherlands and preferred game genre is action.}
\label{tab:ratings_sample}
\begin{tabular}{lccc}
\hline
Game element       & First & Second & Third \\ 
\hline
Acknowledgment     & 0.250 & 0.080 & 0.078 \\
Chance             & 0.000 & 0.144 & 0.059 \\
Competition        & 0.083 & 0.149 & 0.176 \\
Cooperation        & 0.000 & 0.117 & 0.078 \\
Economy            & 0.000 & 0.032 & 0.049 \\
Imposed Choice     & 0.000 & 0.080 & 0.059 \\
Level              & 0.000 & 0.080 & 0.059 \\
Narrative          & 0.000 & 0.043 & 0.020 \\
Novelty            & 0.000 & 0.005 & 0.039 \\
Objectives         & 0.333 & 0.064 & 0.069 \\
Point              & 0.167 & 0.027 & 0.039 \\
Progression        & 0.083 & 0.064 & 0.049 \\
Puzzles            & 0.000 & 0.037 & 0.020 \\
Rarity             & 0.000 & 0.000 & 0.000 \\
Renovation         & 0.000 & 0.005 & 0.010 \\
Reputation         & 0.000 & 0.005 & 0.020 \\
Sensation          & 0.000 & 0.005 & 0.020 \\
Social pressure    & 0.000 & 0.011 & 0.010 \\
Stats              & 0.083 & 0.011 & 0.059 \\
Storytelling       & 0.000 & 0.016 & 0.059 \\
Time Pressure      & 0.000 & 0.027 & 0.029 \\
\hline
\end{tabular}
\end{table}

Aiming to improve the usability of our RS, we reimplemented the CDT generated through the R package \textit{party} \cite{hothorn2006unbiased} in Javascript. Although one could access the R objects, or try to make some external connection to R code from, for instance, a web browser, this process could be laborious and discouraging. On the other hand, Javascript can be easily run in most web browsers, as well as be easily plugged-in into a web site. Furthermore, as decision trees can be represented through a set of if/else statements, the conversion from R objects to Javascript does not require handling complex programming technical challenges. This is another advantage because the procedure of transforming our CDT into a Javascript plugin can be replicated to any other programming language. 

Our RS is freely available (see the appendix), and there are two main use cases in which we believe it can be explored. The first one the main case of automating gamification personalization, in which other systems use it as an external resource/tool. In this case, a gamified system can explore our RS as a plug-in that is consulted to find which game elements should be available for some occasion. To this end, the system would call the plug-in, passing the needed inputs as parameters to receive the ratings of each game element. Then, the system could, for instance, turn on those elements with the highest ratings. This procedure could be iteratively repeated, when the type of the learning activity to be performed changed, for instance. Thus, the RS would aid the system in performing dynamic adaptations \cite{hallifax2019adaptive} of its gamification design according to the user's characteristics and geographic locations as well as the tasks performed. The second case is using our RS as a standalone tool to provide recommendations for one interested in, for instance, personalizing an unplugged gamified environment \cite{toda2019gamify} or to manually define their system gamification.

\section{Discussion}
\label{sec:discussion}

Based on participants' preferences captured though a survey, our findings provided evidence that users' preferences do differ depending on their characteristics, geographic location, and the LAT to be performed (RQ1), as well as we were able to develop an RS that recommends the preferred gamification design for a LAT to be performed by a user with some specific characteristic in a defined geographic location (RQ2). The main contribution of this research is, therefore, providing a free RS for personalized gamification, built upon a state-of-the-art approach, that aids in automating the tailoring of gamification designs by suggesting which game elements to use. This RS is based on the three aspects of personalization: domain, user, and task \cite{liu2017toward}, implemented as the educational domain, demographics and gaming characteristics, and LAT and geographic location, respectively. Additionally, we have shown which context and user characteristics impact their preferences, and which of those are more or less relevant, contributing to expanding and grounding knowledge from previous studies (e.g., \cite{tondello2017elements,denden2017investigation}).

Concerning the results on users' characteristics impacting their preferences, our findings are aligned with the literature. 
Previous studies have shown that, for instance, demographics \cite{toda2019planning,tondello2016gamification} and attributes related to users' gaming habits \cite{denden2017investigation} affect user preference. We corroborate those by providing more empirical evidence that users with different characteristics have different preferences, as well as presenting which of those are more important than others. For instance, we found simple user attributes, such as gender and having researched gamification, are less relevant than gaming-related characteristics (see Table \ref{tab:importance}), which is in line with previous literature suggestions \cite{landers2019defining}. 
Furthermore, it also has been discussed that the task to be performed influences the perceptions of gamified systems' users \cite{liu2017toward}. Following that and within the educational context, suggestions to consider learning activities within the tailoring process of educational systems have emerged \cite{rodrigues2019thinking,hallifax2019adaptive}. Our findings are aligned with those theories as well, showing that users' preferences differ depending on the LAT they expect to perform (see Figure \ref{fig:cdt1} and Section \ref{sec:most_useful_game_elements}). Additionally, we found geographic location to be another relevant factor, finding also consistent with recent literature suggestions \cite{klock2020tailored}.


Concerning the results on users' preferred gamification design for each LAT given their characteristics, we expand the literature by 
i) providing recommendations applicable to any task (by considering its main objective - type) and ii) exploring less studied user characteristics (i.e., demographics and gaming-related) as well as taking into account their geographic location. On one hand, besides not guiding on how to tailor to LAT and geographic location, other personalization approaches (e.g., \cite{oliveira2019tailored,tondello2017elements}) often rely on user profiles \cite{hallifax2019adaptive}. However, as shown by our findings (see Table \ref{tab:importance}), demographics and gaming-related characteristics are relevant as well. On the other hand, despite the recent calls for considering learning activities when personalizing \cite{rodrigues2019thinking,hallifax2019adaptive}, available approaches considering such aspects are yet limited, mainly due to considering only two characteristics (one for from user and the learning activity) \cite{bovermann2020towards,baldeon2016lega}. Although, if multiple aspects are relevant, they all should be considered, as well as their interaction \cite{liu2017toward,buckley2017individualising}. Our research contributes to these concerns, guiding how to personalize gamification to users (i.e., demographics and game-related) and contextual (i.e., LAT and geographic location) aspects simultaneously.

Moreover, this article advances the literature by providing an RS for personalized gamification. In \cite{tondello2017recommender}, a framework for such RS has been proposed, however, the literature still lacks concrete implementations of these systems. On the other hand, recent research has highlighted the need for research to aid in the automation of gamification personalization \cite{klock2020tailored}. This article contributes to this vein by introducing a free RS for personalized gamification that can be both plugged-in gamified systems to automate their personalization process, as well as independently used as a guide for defining personalized gamification designs. As this system is built upon the findings from this article, it implements a state-of-the-art personalization approach, which addresses a couple of literature challenges, namely the need for considering contextual factors along with user information, as well as the interaction between all relevant characteristics (see Section \ref{sec:relatedwork}).

\subsection{Implications} %

There are five main implications of our findings.
First, demographics and game-related characteristics are moderators of user preference that should be prioritized differently. We have shown that these characteristics do affect user preference but that each one's importance differs from one to another. Additionally, those exploring gamification effectiveness might rely on our results to define which data to capture from their samples to further assess whether these characteristics also play a role in other aspects (e.g., motivation or learning from interacting with GES). 

Second, personalization approaches should be expanded beyond the user. We have shown that the game elements people prefer when expecting to perform a LAT differ from what they prefer when expecting to perform another; similarly for users who live in different countries. The implication these findings have is that rather than just thinking on what users generally prefer, aspects of the task that will be performed, as well as the user's geographic location, should be taken into account, as has been recently advocated by multiple researchers \cite{liu2017toward,rodrigues2019thinking,hallifax2019adaptive,klock2020tailored}. 

Third, the interaction between relevant characteristics cannot be ignored. Our results demonstrated that the game elements preferred the most are likely to change when a single characteristic (e.g., country) changes. For example, we demonstrated that the recommended game elements for the same LAT will differ for Brazilian and American users, even if all other characteristics (e.g., gender, weekly playing time, preferred game genre; see Section \ref{tab:recommendation_lat_pgg}) are the same. Thus, confirming the need for tailoring gamification designs not only to the user but also to the context \cite{rodrigues2019thinking,hallifax2019adaptive} as well as considering the interaction between different aspects \cite{liu2017toward,buckley2017individualising}. Hence, the implication is that only one side of the whole is likely not to work in full potential.

Fourth, when designing GES, two people might prefer the same gamification design, but with different priorities. When surveying participants, we asked them to rank the top three game elements that would help them the most in learning activities of a specific type. Hence, gathering data able to inform not only which game elements are the most preferred on each occasion, but also the importance order of the selected elements. Thus, we imply that when relying on our findings to design GES, one should define the emphasis each game element will receive based on users' selection order (see Section \ref{sec:most_useful_game_elements}) because despite different individuals might prefer the same game elements set, they might prefer those with different priorities.

Lastly, putting together our findings and analyses, one can use our RS (see the appendix) to automate gamified systems' personalization process as well as be informed on how to tailor gamification designs of educational systems. Practitioners can exploit our RS to define their systems' gamification designs, as well as researchers can apply its recommendations on their studies to assess the effectiveness of users' preferred designs. To aid those interested in using our RS, we have made it freely available for use and briefly discussed how it can be either incorporated into an existing system as well as using it as a guide. Thus, posing a direct implication on the design and development of GES.

\subsection{Limitations}
\label{sec:limitations}

This section discusses limitations that must be considered when interpreting and applying our findings.

Concerning the survey: It presented a description for each LAT and each game element. We adopted this approach to avoid misinterpretations from those completing the survey, seeking to guarantee answers reliability. However, this increased the time required to complete the survey and possibly contributed to tiring the individuals throughout the process. To address this limitation, we added attention questions, which allowed us to discard inconsistent answers.

Concerning the sample: Although we analyzed data from 361 consistent subjects, their attributes were highly unbalanced in some characteristics (e.g., country). Consequently, recommendations for individuals with a small presence in the dataset (e.g., those reporting a gender other than male and female) are influenced due to the sample size. 

Concerning the RS: Although we selected the revision of Bloom's taxonomy due to its relevance within the education context, the lack of a systematic selection process also limits our findings in terms of how our study interprets LAT. Also, as our recommendations are based on averages, it might be that it will not work for some users. Lastly, although our RS is a ready-to-use resource, it is a plug-in in its initial version that can be further enhanced to improve, for instance, its compatibility with other systems, as well as its presentation for independent use.

Concerning recommendations' effectiveness: This was an exploratory, preference-based research, following a methodology commonly adopted by related research. Consequently, as these previous research, we cannot ensure that personalizing to users' preferences will be effective. Nevertheless, given the number of game elements to be considered (21) as well as LAT (six), thousands of combinations would have to be tested in user studies, which is unfeasible. Our survey-based study addresses this limitation by presenting a valuable first step in suggesting which game elements to use for specific conditions and providing guidance for future studies to test our preference-based recommendations. Nevertheless, this limitation suggests the need for studies to design ways of testing the level of preference between all game elements and all LAT. 

\section{Final Remarks}
\label{sec:finalremarks}

Personalization emerged as an alternative to improve gamification effectiveness. Most studies in this vein exploit user profiles to tailor the gamified designs. Hence, they ignore the fact that, besides the user, the tasks and domain also play a significant role in gamification's success. Additionally, studies often do not consider the interaction between multiple relevant characteristics, neither offer concrete resources to help in automating gamification personalization. To address these gaps, this paper introduced a preference-based RS that suggests game elements tailored to the user (demographics and game-related) and the context (LAT - tasks - and geographic location), focused on the educational domain. This RS considers the interaction between its inputs and is freely available for anyone to use it.

Our contributions are twofold. First, we provided practitioners with a ready-to-use resource able to guide them on how to design GES that are tailored to users' characteristics, as well as geographic location, according to the tasks they are up to perform. Second, we expanded the literature on how to tailor gamification designs to any learning activity (based on its type) by presenting recommendations that might be empirically tested in future research, providing empirical evidence on which demographics and game-related user characteristics impact their preferences, as well as whether one is more important than another, and supporting literature suggestions by showing that LAT and geographic location do affect user preference.

As future studies, we mainly recommend validating the effectiveness of our RS recommendations (e.g., ability to improve user motivation, flow, academic performance, or learning gains), compared to one-size-fits-all and other personalization methods, to identify whether personalizing to users' preferences will positively impact them as expected. Another line of future research is improving the RS so that is can be used as, for instance, a service to mitigate compatibility problems as well as the need for manually adding the code to the project. Additionally, future studies might tackle the limitation of not assessing the match between all game elements and all LAT from our methodology, which might be accomplished in steps (e.g., assessing one LAT per experiment) to cope with the complexity of testing all at once, as we previously discussed.  

\section*{Appendix} Appendixes are available at: \url{shorturl.at/aguQT}.

\section*{Acknowledgments}

The authors would like to thank the funding provided by CNPq, CAPES, and FAPESP (Projects: 2018/07688-1; 2018/15917-0; 2016/02765-2; 2018/11180-3).



\begin{IEEEbiography}[{\includegraphics[,width=1in,height=1.25in,clip,keepaspectratio]{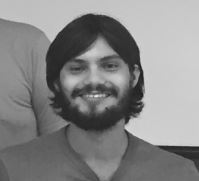}}]{Luiz Rodrigues}
Luiz Rodrigues is a Ph.D. candidate at the University of São Paulo. His research focuses on understanding and improving gamification effectiveness within the context of educational systems. His main research interests are gamification, personalization, user modeling, and procedural content generation.
\end{IEEEbiography}

\begin{IEEEbiography}[{\includegraphics[width=1in,height=1.25in,clip,keepaspectratio]{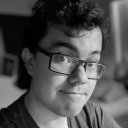}}]{Armando M. Toda}
Armando Toda is, currently, a PhD. candidate at University of Sao Paulo working with gamification design in educational contexts. Other main areas of research are concerned with: data mining, serious games, evaluation and instrument design.
\end{IEEEbiography}

\begin{IEEEbiography}[{\includegraphics[width=1in,height=1.25in,clip,keepaspectratio]{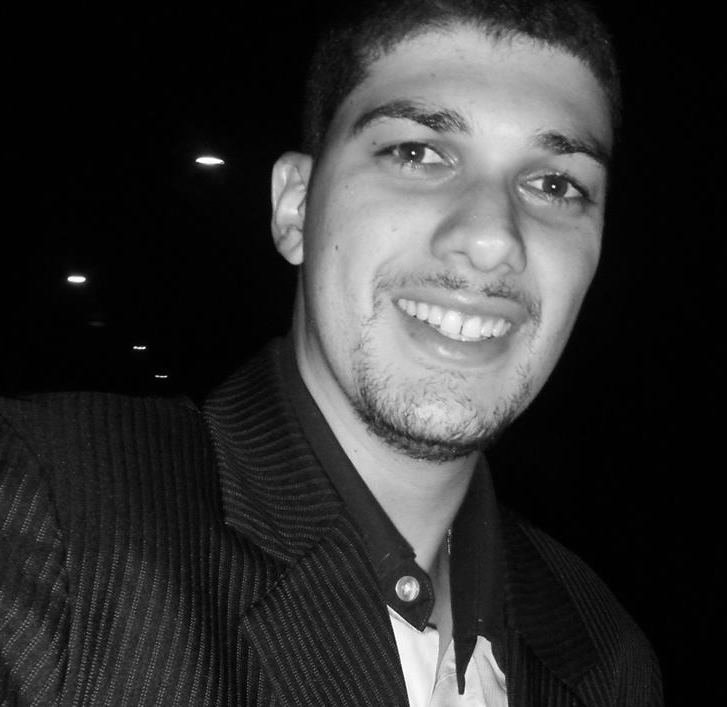}}]{Wilk Oliveira}
Wilk Oliveira is a Ph.D. candidate at the University of São Paulo and MSc. in Computer Science for the Federal University of Alagoas with exchange programs at the University of Saskatchewan. He was an assistant professor at the University of São Paulo and a guest lecturer at the Tiradentes University Center. His main topics include Artificial Intelligence in Education, Human-computer Interaction, and Gamification.
\end{IEEEbiography}

\begin{IEEEbiography}[{\includegraphics[width=1in,height=1.25in,clip,keepaspectratio]{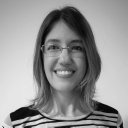}}]{Paula T. Palomino}
Paula Palomino is a PhD. candidate at University of São Paulo, researching and developing a gamification content-based framework for educational purposes. Other areas of study include Games, HCI and Cyberculture.
\end{IEEEbiography}

\begin{IEEEbiography}[{\includegraphics[width=1in,height=1.25in,clip,keepaspectratio]{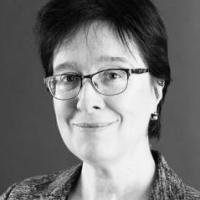}}]{Julita Vassileva}
Julita Vassileva is Professor at the Department of Computer Science, University of Saskatchewan. Julita does research in Social Computing, Persuasive Technologies, Personalization, User Modeling, AI in Education, Peer-to-Peer systems, Trust and Reputation Mechanisms, and Human-computer Interaction. She is currently working on Personalized Persuasive Systems, Learning Data-Driven Persuasion, and Distributed Ledgers for Storing and Sharing Personal Data.
\end{IEEEbiography}

\begin{IEEEbiography}[{\includegraphics[width=1in,height=1.25in,clip,keepaspectratio]{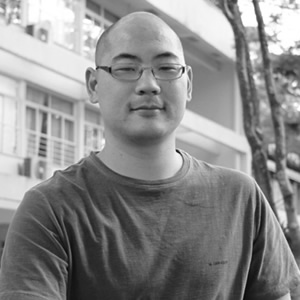}}]{Seiji Isotani}
Seiji Isotani is a full professor at the University of São Paulo (USP). His research topics include Artificial Intelligence in Education, Intelligent Tutoring Systems, Computer-Supported Collaborative Learning (CSCL), Gamification, Ontology Engineering and Linked Open Data. He has published more than a hundred scientific papers, has more than 3000 citations and, according to Google Scholar, is among the 15 top researchers in gamification
\end{IEEEbiography}

\end{document}